\documentclass[twocolumn]{aa}
\usepackage{graphicx}
\usepackage{aalongtable}
\usepackage{chemarr}
\usepackage{amssymb} 
\usepackage{amsmath} 
\usepackage{mathrsfs}
\usepackage{txfonts}
\usepackage{natbib}
\usepackage{color} %--temporarily inserted -- will be removed 
\bibpunct{(}{)}{;}{a}{}{,}
\defcitealias{morisawa}{MFK}
\begin{document}
  \title{Modeling  the ortho-to-para abundance ratio of \\
	 cyclic C$_3$H$_2$ in cold dense cores}

  \author{I. H. Park\inst{1}, V. Wakelam\inst{2}, \and E. Herbst\inst{3}}
  \offprints{E. Herbst}
  \institute{Chemical Physics Program, The Ohio State University,
	     Columbus, OH 43210 USA\\
             \email{ihpark@mps.ohio-state.edu}
             \and
	     Department of Physics, The Ohio State University,
	     Columbus, OH 43210 USA\\
             \email{wakelam@mps.ohio-state.edu}
             \and
             Departments of Physics, Chemistry and Astronomy,
             The Ohio State University, Columbus, OH 43210 USA\\
             \email{herbst@mps.ohio-state.edu}
            }
  \date{Received 2005; accepted 2005}
  \abstract
  {}{We report a detailed attempt to model the ortho-to-para abundance 
	ratio of $c$-C$_3$H$_2$ so as to reproduce observed values in the cores 
	of the well-known source TMC-1. 
        According to observations, the ortho-to-para ratios vary, within large uncertainties, 
        from a low of 
        near unity to a high of approximately three depending on the core. }
        {We used the  osu.2003 network of gas-phase chemical reactions
        augmented by reactions that specifically consider the formation, 
        depletion, and interconversion of the $ortho$ and $para$ forms of the 
        $c$-C$_{3}$H$_{2}$ and its precursor ion $c$-C$_{3}$H$_{3}^{+}$. 
        We investigated the sensitivity of the  calculated ortho-to-para ratio 
        for $c$-C$_{3}$H$_{2}$ to a large number of factors. }
        {For the less evolved cores C, CP, and D, we had no difficulty 
        reproducing the observed ortho-to-para ratios of 1-2.  
        In order to reproduce observed ortho-to-para ratios of near three, 
        observed for the evolved cores A and B, it was necessary to include 
        rapid ion-catalyzed interconversion processes. } {}

  \authorrunning{Park, Wakelam, \& Herbst}
  \titlerunning{Modeling the $o/p$ C$_{3}$H$_{2}$ ratio in TMC-1}
  \maketitle
%
%------------------------------------------------------------------------------
% 1. 
\section{Introduction}
%
%------------------------------------------------------------------------------
% 
   % Studies about C3H2 interstellar molecules
   %==========================================
   Cyclopropenylidene, $c$-C$_3$H$_2$, is a widely-distributed abundant 
    organic ring molecule in the interstellar medium \citep{Ref:c3h2-10,mat85,Ref:c3h2-12}.
   Its ``linear'' isomer, propadienylidene ($l$-C$_3$H$_2$), has also been 
   observed in a number of sources, albeit at lower 
   abundance \citep{Ref:c3h2-7,Ref:c3h2-5,teyssier,Ref:c3h2-6,Ref:c3h2-66,Ref:c3h2-8}.
   For each isomer, two equivalent H nuclei, of spin 1/2, 
   couple to generate $ortho$ (nuclear spin = 1) and $para$ (nuclear spin = 0)  
   species with spin statistical weights of 3 and 1, respectively. 
      The nuclear spin wave functions of the $ortho$ states are symmetric to 
   exchange of the protons whereas those of the $para$ states are anti-symmetric. 
   Since the Pauli exclusion principle requires total wave functions to be
   antisymmetric to exchange of protons, the symmetry of the rotational wave
   function of the molecule to exchange must be antisymmetric for $ortho$ spin 
   states and symmetric for $para$ spin states if the ground electronic state 
   is symmetric to exchange. The dependence of the exchange symmetry of the
   rotational state on the rotational quantum numbers depends on the structure
   of the molecule and on the symmetry 
   of the electronic state \citep{townes}. Moreover, the conversion of molecular 
   species from rotational levels with $ortho$ spin functions to those with 
   $para$ spin functions does not occur  efficiently if at all by radiative or 
   non-reactive collisional mechanisms in the gas, so that, in the absence of 
   reaction (or strong binding with a surface),  the sets of rotational levels 
   with $ortho$ and $para$ spin functions can be regarded as distinct species.

   %  Observations of ortho and para forms of C3H2
   %=====================================================================
   The cyclic isomer of C$_{3}$H$_{2}$ has been observed in the interstellar medium in transitions
   belonging to both spin states:  for $o$-C$_3$H$_2$, the observed lines 
   include the $b$-type transitions $1_{10}$-1$_{01}$, 2$_{12}$-1$_{01}$, 
   3$_{12}$-3$_{03}$, 3$_{21}$-3$_{12}$ and $3_{30}$-3$_{21}$ at 18.3, 85.3, 
   83.0, 44.1, and 27.1 GHz respectively, whereas for $p$-C$_3$H$_2$, the 
   observed $b$-type transitions include 2$_{02}$-1$_{11}$, 2$_{11}$-2$_{02}$,
   2$_{20}$-2$_{11}$ and 3$_{22}$-3$_{13}$ lines at 82.1, 46.8, 21.6, and 84.7
   GHz respectively 
   \citep{Ref:c3h2-7,mat85, Ref:c3h2-10,Ref:opC3H2,Ref:BR-madden,morisawa}.  
   With this many transitions, ortho-to-para ratios of reasonably high accuracy 
   can be obtained, allowing a detailed understanding of the chemical processes 
   involved and possibly even the history of the source.
   
   %  Laboratory work on ortho, para isomers 
   %=====================================================================
   In the terrestrial laboratory one often refers to two-types of mixtures for 
   $ortho$ and $para$ species involving two protons: the common $normal$ 
   variety, in which the ortho-to-para ratio is three, and the $equilibrium$ 
   variety, in which the ortho-to-para ratio reflects thermal equilibrium.  In the interstellar medium, however, 
    the ortho-to-para  ratio at low temperatures is 
   probably governed by kinetic rather than thermodynamical considerations, including processes on grains, and
   is likely to be time-dependent and may even reflect the original circumstances of the molecule formation \citep{tine,Ref:opH2-2,Ref:opH2-3}. 
   In fact, detailed estimates of the steady-state interstellar ortho-to-para H$_2$ ratio at 
   10 K show that it is approximately 10$^{-3}$
   \citep{opH2-ratio,opH2-non-thermal,cooling-H2}, far higher than the equilibrium value of $3 \times 10^{-7}$.  

   Molecular hydrogen is unique in showing such a strong deviation between thermal 
   and normal values for the $o/p$ ratio at temperatures in the vicinity of 10 K.  
   For heavier species, this deviation is much smaller because more levels of each spin modification are thermally populated so that values of $\approx 3.0$ for the thermal ratio pertain to quite 
   low temperatures. For H$_{2}$CO, \citep{Ref:H2CO-1,Ref:H2CO-2},  this 
 ratio is reached at temperatures as low as $\approx$ 15 K, while for
  $c$-C$_{3}$H$_{2}$, the 
    ratio is reached under thermal conditions at temperatures under 5 K 
   \citep{Ref:opC3H2}. Nevertheless, the actual $o/p$ ratio in cold dense interstellar sources is determined by kinetic 
   considerations and may well be significantly under three at 10 K
   clouds for heavy species with two equivalent protons, as detected for the species 
   H$_2$CO, H$_2$CS, and H$_2$CCO \citep{ Ref:H2CO-1,minh,ohishi}.
  
   % Studies about C3H2 interstellar molecules
   %==========================================
   Although the ortho-to-para abundance ratio of $c$-C$_3$H$_2$ had been studied 
   observationally in TMC-1 by \cite{Ref:BR-madden} and \cite{Ref:opC3H2} and 
   in L1527 by \citet{Ref:opC3H2}, this work is motivated by the recent 
   observational work of \citet{morisawa} (hereafter, MFK), who measured 
   the ratio in a more comprehensive manner for the six different cores  within 
   the TMC-1 ridge: A, B, C, D,  CP, and E. Their measured ratios range from 
   1.4$\pm$0.7 (TMC-1D and CP) to 3.0$\pm$1.5 (TMC-1A and B). 
   Note that Core D sometimes refers to some substructure near the CP 
   condensation, in which case, it can be referred to as CP-b 
   \citep{TMC-1-CP-b}, but  other authors have used Core D to be the same as 
   Core CP \citep{ Ref:ism-3-2}, which is the very well-studied core where the 
   abundances of complex carbon chains peak \citep{smith04}. In addition, TMC1-B is the
   so-called ammonia peak \citep{age-howe}.  The current understanding is that some of 
   the cores (e.g. Core B) are more evolved than others (e.g. Core CP) although the 
   cause of the differential evolution is not fully understood 
   \citep{Ref:TMC-models,age-howe,markwick}. 
   
   Table \ref{op-obs} shows the observed 
   $o/p$ abundance ratios of cyclic $c$-C$_{3}$H$_{2}$ in the six cores 
   \citepalias{morisawa} along with  measured fractional abundances (X) of 
   $o$-C$_{3}$H$_{2}$ with respect to H$_{2}$  and estimated ages of the cores. 
   Although the determination of cloud ages is still contentious, there is a sufficient consensus
   concerning the chemical indicators of age to use the results in this paper.
   It can be seen that, as noted by \citetalias{morisawa}, the ortho-to-para ratio 
   seems to be generally correlated with the evolution of the core; i.e., for all cores other than Core E,  the older the core 
   chemically, the higher the ortho-to-para ratio within large uncertainties, especially for the higher ortho-to-para ratios.
   %%%%%%%%%%%%%%%%%%%
   \begin{table*}     
	\caption[]{Some c-C$_{3}$H$_{2}$ observations of the cores in the TMC-1 ridge}     
	$$     
	\begin{array}{p{0.225\linewidth}p{0.225\linewidth}p{0.225\linewidth}p{0.225\linewidth}} 
	\hline        
%	\noalign{\smallskip}  
%	\noalign      
	Source     &  Observed $o/p$ Ratio$^b$       		& Observed X($o$-C$_3$H$_2$) $\times\ 10^{-10}$ $^{b}$  & Estimated Age (yr) \\\noalign{\smallskip}        
	\hline\noalign{\smallskip}         
	{\it Cores with high $o/p$ ratio\ $\cdots$}   		& 	        &              	& \\\noalign{\smallskip}        
	TMC-1A      & 3.0$^{+2.2}_{-1.3}$    	    		& 7.5$^{+12}_{-5.4}$	& $\sim$ 10$^6$ $^b$ \\\noalign{\smallskip}        
	TMC-1B      & 3.0$^{+1.7}_{-1.1}$; 3.0$\pm$0.16$^{c}$   & 7.9$^{+12}_{-5.4}$	& $\sim$ 10$^6$-10$^7$ $^h$ \\\noalign{\smallskip}        
	TMC-1E      & 2.8$^{+2.1}_{-1.2}$   	 	    	& 4.6$^{+7.7}_{-3.4}$	& $\sim$ 10$^4$-10$^5$ $^{a,b}$ \\\noalign{\smallskip}         
	{\it Cores with low $o/p$ ratio\ $\cdots$}    	&       &              	& \\\noalign{\smallskip} 
        TMC-1C      & 1.8$^{+0.9}_{-0.6}$; 2.4$\pm$0.1$^{c}$    & 6.0$^{+8.4}_{-4.0}$; 13 $^{e}$	& $\sim$ 10$^5$ $^b$ \\  \noalign{\smallskip}        
        TMC-1CP     & 1.4$^{+0.9}_{-0.5}$; 2.56$\pm$0.31$^{d}$  & 16$^{+32}_{-13}$; 57 $^{d}$ 	& $\sim$ 10$^5$ $^{f,g}$; $6 \times 10^4$ $^h$ \\  \noalign{\smallskip}        
        TMC-1D$^a$  & 1.4$^{+1.1}_{-0.7}$                      	& 8.3$^{+19}_{-6.9}$ 		 		& $6 \times 10^4$ $^h$ \\ \noalign{\smallskip}         
	\hline      
	\end{array}      
	$$      
	\label{op-obs} 
     	\begin{list}{}{}	 
	\item  $^a$ \citet{TMC-1-CP-b}. Core D is Core CP-b according to this source. 
 	\item  $^b$ \citetalias{morisawa} unless noted. The fractional abundances X of $o$-
	C$_{3}$H$_{2}$ are with respect to H$_{2}$.  Based on the MFK logarithmic observables,
		we carried out slightly different error calculations, because here
		$\Delta$ logX is not much less than 1, so that two 1$\sigma$ random errors $\Delta{X}_{-}\napprox\Delta{X}_{+}$   should be calculated \citep{wakelam} .		
 	\item  $^c$ \citet{Ref:opC3H2}
 	\item  $^d$ \cite{madphd} 
	\item  $^e$ \citet{cox-c3h2}
	\item  $^f$ \citet{age-D}
	\item  $^g$ \citet{smith04}	
	\item  $^h$ \citet{Ref:ism-3-2} 
     	\end{list} 
 	\end{table*}
   %%%%%%%%%%%%%%%%%%%
   
   \citetalias{morisawa} used a simple model of a small number of gas-phase chemical 
   reactions to understand their measured abundance ratios in terms of the evolution 
   of the cores. They showed that the ortho-para chemistry of $c$-C$_{3}$H$_{2}$ is 
   rather simple compared with other heavy molecules that have been studied, so that 
   a complete understanding of the $o/p$ abundance ratio is possibly obtainable. Our 
   detailed model is based on their simple one, with, as discussed below, the 
   important reactions for the formation and depletion of the $ortho$ and $para$ 
   forms of $c$-C$_{3}$H$_{2}$ and its protonated ion precursor embedded in a large 
   chemical network.\\

   In this paper, we report the results of our detailed gas-phase chemical model of 
   the ortho-to-para abundance ratio of $c$-C$_{3}$H$_{2}$.  
   The remainder of this paper is divided as follows. 
   In Section 2 we discuss the details of our gas-phase chemical model, 
   while in Section 3, we illustrate our results for the ortho-to-para ratio of 
   $c$-C$_{3}$H$_{2}$ as a function of time. We also compare our results with the 
   model of \citetalias{morisawa}.
   As will be seen, our standard results are not capable of producing $o/p$ ratios greater than
   2.1, so we consider some additional direct chemical processes that can convert 
   the $para$ form of the ring molecule to its $ortho$ form. With these processes, 
   we are able to account for the possibly high $o/p$ ratio in the more evolved cores. 
   We then subject our results to an uncertainty analysis. Section 4, the discussion, 
   concludes the paper.

%
%------------------------------------------------------------------------------
% 2. 
\section{The model}
%
%------------------------------------------------------------------------------
%
   %%%%%%%%%%%%%%   
   %%%%%%%%%%%%%%   
   %=================
   % Reaction network
   %=================
   The chemical reaction network used for this work is an updated version of the 
   {\sc osu.2003} gas-phase network  \citep{smith04}.  Several versions of this 
   network are now available at the URL \texttt{http://{\allowbreak}www.{\allowbreak}physics.{\allowbreak}ohio-state.{\allowbreak}edu/{\allowbreak}$\sim$eric{\allowbreak}/research.html}. 
   For this work, we added several other modifications. First, and most importantly,
   we added reactions to account for the chemistry of both forms of $c$-C$_{3}$H$_{2}$ 
   and its protonated precursor $c$-C$_{3}$H$_{3}^{+}$ (see below). Note that this 
   ion has three-fold symmetry so that its protons are all indistinguishable: 
   the $ortho$ form has a total nuclear spin of 3/2, while the two $para$ forms \citep{Ref:oka04} each 
   have a total nuclear spin of 1/2,  although only one of these exists for a given rotational state \citep{townes}. Secondly, we added new experimental information 
   on the neutral products and total rate coefficients for dissociative recombination 
   reactions involving several hydrocarbon ions \citep{CnHm1,CnHm2,CnHm4,CnHm3,CnHm5}.
   Thirdly, we ignored the existence of the linear isomer of lower abundance - 
   $l$-C$_{3}$H$_{2}$ -  to simplify the problem for $c$-C$_{3}$H$_{2}$. 
   % 
   %---------------------------------------------------------------------------
   \subsection{Formation and destruction of $c$-C$_3$H$_2$}
   %---------------------------------------------------------------------------
   %
   The formation of $c$-C$_3$H$_2$ begins with the radiative association of 
   C$_3$H$^+$ with H$_2$ to produce the precursor ion $c$-C$_3$H$_3^{+}$. This 
   association is not the only reaction these two reactants undergo; the H-atom 
   transfer channel to form $c$-C$_{3}$H$_{2}^{+}$ + H may be competitive, although 
   it is unclear whether or not the reaction is endothermic 
   \citep{Ref:c3h2-4,Ref:BR-maluendes}. In the {\sc osu}.2003 \citep{smith04}
   and {\sc rate99} networks \citep{teuff} the H-atom transfer is assumed to be 
   a minor channel at 10 K. Once the cyclic C$_{3}$H$_{3}^{+}$ ion is formed, 
   its major destruction is by dissociative recombination, one channel of which 
   leads to $c$-C$_{3}$H$_{2}$:
   \begin{eqnarray}\label{rxn:init}
      \mbox{C}_3\mbox{H}^+     +  \mbox{H}_2   	   
	\to &  c\mbox{-C}_3\mbox{H}_3^+  +       h\nu     & (\mbox{major}) \\
	\to &  c\mbox{-C}_3\mbox{H}_2^+  +       \mbox{H} & (\mbox{minor})\\
      c\mbox{-C}_3\mbox{H}_3^+ +  \mbox{e}\label{BR-DR1} 	   
	\to &  c\mbox{-C}_3\mbox{H}_2    +       \mbox{H} & \\\label{BR-DR2}
	\to &  c\mbox{-C}_3\mbox{H}        +       \mbox{...} & 
   \end{eqnarray}
  where the $\dots$ in reaction (\ref{BR-DR2}) refer to the fact that the additional products H + H or H$_2$
   have not been explicitly determined by experiment.
   According to the recent experiment performed by \citet{CnHm6} on the dissociative 
   recombination of C$_3$H$_3^+$, product channels containing molecules with three 
   carbon atoms account for 90.7\% of the products. The experiment was not able to 
   differentiate among the C$_{3}$H$_{m}$ species, nor to determine if linear or 
   cyclic species were present, so the branching ratio for the important products 
   $c$-C$_{3}$H$_{2}$ + H remains a free parameter.  

   Once produced, $c$-C$_{3}$H$_{2}$ is destroyed by reactions with a variety of ions 
   while its precursor ion can be destroyed by neutral species in addition to 
   dissociative recombination:	
   \begin{eqnarray}
      c\mbox{-C}_3\mbox{H}_2 + \mbox{[H}_3^+ \mbox{,\ } \mbox{HCO}^+ \mbox{,\ } \mbox{H}_3\mbox{O}^+\mbox{]} & \to & c\mbox{-C}_3\mbox{H}_3^+ + \mbox{X} \\	 	
      c\mbox{-C}_3\mbox{H}_2 + \mbox{[C}^+\mbox{,\ }\mbox{S}^+\mbox{,\ }\mbox{Si}^+\mbox{]}  & \to & \mbox{products}\\
            c\mbox{-C}_3\mbox{H}_3^+ + \mbox{[C,\ O,\ S,\ Si]}  & \to & \mbox{products}
   \end{eqnarray}
   Note that from now on, the cyclic designation $c$- will be understood and not 
   written explicitly.  
   \begin{figure}
   \includegraphics[width=1\linewidth]{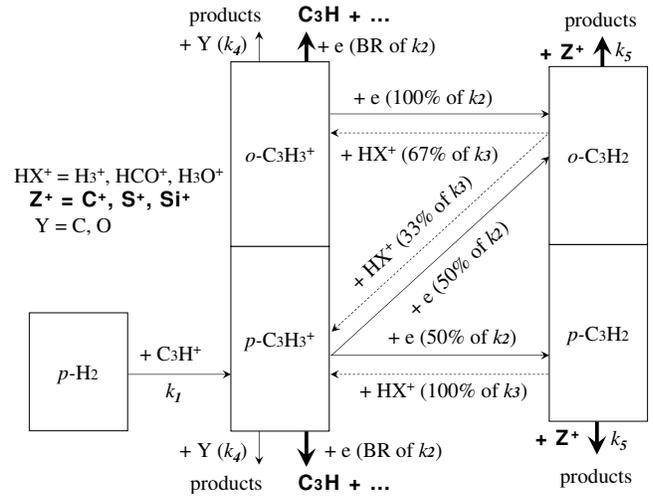}
   \caption{Dominant reaction diagram for the standard models. Some
   processes not considered by  \citetalias{morisawa} are indicated by boldface.}
   \label{dominant-rxn}
   \end{figure}
   
   One can see that in the absence of C$_{3}$H$_{2}$ reactions involving ions other than 
   protonating ones, of ion-atom reactions involving C$_{3}$H$_{3}^{+}$, and of 
   dissociative recombination of this ion to form any products other than 
   C$_{3}$H$_{2}$, there is a quasi-cycle or loop in which once the protonated ion 
   C$_{3}$H$_{3}^{+}$ is formed, dissociative recombination and protonation serve 
   only to  convert these species into one another. As these processes 
   occur, ortho-para conversion also occurs, as we shall now discuss. 
   %
   %---------------------------------------------------------------------------
   \subsection{Para-to-ortho conversion:
   	       standard and alternative models} \label{SS:2.2}
   %---------------------------------------------------------------------------
   %
   The dominant set of reactions involving the ortho/para chemistry of 
   C$_{3}$H$_{2}$ in our {\em standard models}, based on the model of 
   \citetalias{morisawa}, is listed in Table \ref{op-raxn} along with additional 
   reactions for two {\em alternative models}. The important reactions in the 
   standard approach are shown in Fig. \ref{dominant-rxn} where the rate 
   coefficients and their numerical subscripts are defined in the table.  
   In Table \ref{op-raxn}, the reactions are on the far left, with the rate 
   coefficients and branching fractions next to them. The branching fractions 
   are written in terms of the rate coefficients, which are numbered according 
   to their position in the table. Reactions in the simple model 
   of \citetalias{morisawa} are indicated by a dark circle.
  
   The basic structure of the reaction scheme commences with the production of 
   $p$-C$_{3}$H$_{3}^{+}$ by  radiative association because the H$_{2}$ 
   reactant, as discussed in the Introduction, is overwhelmingly $para$ in 
   nature for cold clouds. Thus the zero nuclear spin of H$_{2}$ and the spin 
   of the individual proton in C$_{3}$H$^{+}$ can combine only to form a 
   resultant nuclear spin of 1/2, which corresponds to a $para$ form of 
   C$_{3}$H$_{3}^{+}$. Once $p$-C$_{3}$H$_{3}^{+}$ is formed, it undergoes 
   dissociative recombination to form both $p$- and $o$- forms of
   C$_{3}$H$_{2}$ (as well as other neutral products). The C$_{3}$H$_{2}$ 
   species can then undergo protonation reactions with a variety of ions 
   designated HX$^{+}$; these processes lead to both $ortho$ and $para$ 
   versions of the protonated ion. As the loop reactions continue, sizable 
   proportions of $o$-C$_{3}$H$_{2}$ are built up. In competition with the 
   production of $ortho$ species within the loop, other reactions serve to 
   force the system to exit the loop, including (i) non-protonating 
   ion-molecule reactions with C$_{3}$H$_{2}$, (ii) dissociative recombination
   of C$_{3}$H$_{3}^{+}$ to form products other than C$_{3}$H$_{2}$, (iii) 
   ion-atom reactions involving C$_{3}$H$_{3}^{+}$.  Of course, in a complex 
   model, some of the products formed outside of the loop can eventually lead
   to reactions such that the products are once again C$_{3}$H$_{2}$ and its
   protonated ion. In addition to the cycle of protonation and dissociative
   recombination, there may also be direct conversion processes; such processes will be discussed later but 
   can be seen in Table \ref{op-raxn} as  part of our alternative models.
  
   In order to determine the selection rules and branching ratios for 
   ortho/para products in protonation and dissociative recombination, we have
   used the approach of \citet{Ref:oka04}, in which angular momentum algebra is
   applied to the addition of the nuclear spins of the reactants to determine 
   the spin angular momenta of the intermediate complex, which then dissociates 
   into product states of given spin angular momentum. In the language of the 
   rotation group, if one considers the two reactants to have proton nuclear 
   spins of $I_{1}$ and $I_{2}$, the combined spin state can be written as 
   $\mathscr{D}_{I_1} \otimes \mathscr{D}_{I_2}$.  
   This state leads to the intermediate state 
   $\mathscr{D}_{I_1+I_2} \oplus \mathscr{D}_{I_1+I_2-1} 
    \oplus \cdots \oplus\mathscr{D}_{|I_1-I_2|}$,
   which means that the complex can have total proton spin angular moment 
   ranging from $I_{1}$ + $I_{2}$ down to $|I_{1} - I_{2}|$.  Each of these 
   angular momenta can then dissociate into product spin states given by 
   angular momentum rules \citep{Ref:oka04}. The approach of Oka is most valid
   for processes in which many rotational states of the products can be 
   produced. These processes can be exothermic reactions or even thermoneutral 
   reactions if the energy is sufficiently high. In this limit, one need not 
   worry much about the rotational quantum number $J$, since many different $J$
   states are produced.  For low-temperature thermoneutral processes, on the 
   other hand, where a small number of $J$ states may be produced, the role of 
   the individual $J$ states needs to be taken into account in a comprehensive 
   quasi-phase-space theoretical approach \citep{herb82,Ref:opH2-5,gerl02}. 
   Although the approach of \citet{Ref:oka04} lends itself naturally to the 
   idea of a reaction intermediate in which all protons are indistinguishable
   from one another and can be facilely exchanged during the life of the 
   complex, it is also possible to remove this assumption. In fact, in our use 
   of the treatment for the ion H$_{3}^{+}$, we assume that the protonation 
   reactions occur by simple transfer of a proton rather than formation of a 
   complex consisting of C$_{3}$H$_{2}$ and H$_{3}^{+}$. Our branching 
   fractions are the same as those used by \citetalias{morisawa}, who derived
   theirs with the approach of \citet{quack}, which is equivalent to that of 
   \citet{Ref:oka04}.
  
   In our standard models, we do not consider direct (ion-catalyzed) 
   interconversion processes between the $ortho$ and $para$ forms of C$_{3}$H$_{2}$ 
   because competitive reactions with abundant ions that might catalyze such a 
   conversion also occur. For example, the reaction of C$_{3}$H$_{2}$ with protons 
   is likely to produce exothermic products such as C$_{3}$H$^{+}$ + H$_{2}$ rather
   than undergo the exchange of protons needed to cause ortho/para change while the 
   reactions with protonating ions such as HCO$^{+}$ and H$_{3}^{+}$ are likely to 
   undergo protonation. Nevertheless, we find, as will be shown Section \ref{SS:3.2}, 
   that in the absence of such interconversion/exchange reactions, the calculated 
   ortho-to-para ratio for C$_{3}$H$_{2}$ cannot be large enough to explain observed 
   values near 3.0.  
   So, in our  alternative models we also consider interconversion of $o$- and 
   $p$-C$_{3}$H$_{2}$ by ion-catalyzed exchange processes with the four ions H$^{+}$, 
   H$_{3}^{+}$, H$_{3}$O$^{+}$ and HCO$^{+}$, the latter three occasionally being
   designated by HX$^{+}$.  For these processes,  the approach of \citet{Ref:oka04} 
 leads to the detailed balance expression in the limit that $kT$ 
   exceeds the ortho-para energy difference:	 
   \begin{equation} 
      \frac{k(p \to o)}{k(o \to p)}  = \frac{k_{-i}}{k_{i}}  =  3,
   \end{equation}
   where $i$ is an index of the four pairs of interconversion processes considered.
   We simplify the forward and backward processes such that the reactants produce
   a complex with a given rate coefficient $k = k_{i} + k_{-i}$; the complex then 
   dissociates into $o$-C$_{3}$H$_{2}$ on 3/4 of the collisions and 
   $p$-C$_{3}$H$_{2}$ on 1/4 of the collisions regardless of the initial reactants. 
   In the theory of \citet{Ref:oka04}, the relevant formulae are 
   $k(o \rightarrow p)$ = 1/6  $k$ and $k(p \rightarrow o)$ = 1/2  $k$ for isomerization reactions  
   involving a proton.   For the unknown rate coefficients 
   $k_{i} + k_{-i}$, we choose a standard value equal to the ion-dipole collision 
   rate.  To the best of our knowledge, there is no evidence for or against the
   supposition that these particular ortho/para exchange reactions are not 
   competitive with reaction. Perhaps the strongest evidence that such processes 
   occur is discussed in \citet{cordon}, where the exchange mechanism via the complex
   H$_{5}^{+}$ is measured to occur in the reaction between H$_{2}$ and H$_{3}^{+}$ 
   in competition with a simple proton hop.       
   %
   %---------------------------------------------------------------------------
   \subsection{Parameters and models: 
	     default \& sets of models 1-6}
   %---------------------------------------------------------------------------
   %
   %%%%%%%%%%%%%
   \setcounter{table}{2}
   \begin{table}
      \caption{Default initial conditions}
      \label{Table:default}
      \centering
      \begin{tabular}{lccc}
        \hline
        \noalign{\smallskip} 
      	\multicolumn{2}{c}{Low-metal abundance} & Other parameters \\
        \noalign{\smallskip} \hline
        \noalign{\smallskip}
	He       & 6.00(-2) 			 & 		\\ 
	N    	 & 2.14(-5)     		 &           	\\
	O     	 & 1.76(-4)	   		 &           	\\
	H$_2$ 	 & 5.00(-1)	  		 & Temperature = 10 K \\ 
	C$^+$    & 7.30(-5)	  		 & n$_H$ = 2$\times$10$^{4}$ cm$^{-3}$	\\
	S$^+$    & 8.00(-8)	  		 & $\zeta$ = 1.3$\times$10$^{-17}$ s$^{-1}$	\\
	Si$^+$   & 8.00(-9)	  		 & C/O ratio = 0.41   	\\
	Fe$^+$   & 3.00(-9)	  		 & $A_v$ = 10\ mag \\ 
	Na$^+$   & 2.00(-9)	  		 &           	\\
	Mg$^+$   & 7.00(-9)	  		 & C$_3$H$_3^+$ + e $\xrightarrow{0.5\ k_2}$ C$_3$H$_2$ + H\\
	P$^+$    & 3.00(-9)	  		 & \hspace{36pt}$\xrightarrow {0.5\ k_2}$ C$_3$H + ...\\
	Cl$^+$   & 4.00(-9)	  		 &              \\ 
	e 	 & 7.3107(-5)	  		 &           	\\
        \noalign{\smallskip}
	\hline
        \noalign{\smallskip}
      \end{tabular}
      \begin{list}{}{}
	 \item  Note : a(-b) = a $\times$ 10$^{-b}$, with respect to H abundance 
      \end{list}
   \end{table}
   %%%%%%%%%%%%%
   %%%%%%%%%%%%%%
   \setcounter{table}{3}
   \begin{table}
      \caption[]{Summary of variations of standard model parameters}
      $$
      \begin{array}{p{0.45\linewidth}p{0.45\linewidth}}
	 \hline
         \noalign{\smallskip}
         Parameters Varied &  Range \\
         \noalign{\smallskip}
         \hline
         \noalign{\smallskip}
         1) C/O ratio  				& 0.21 -- 1.20\\
         2) Metal depletion			& low-metal -- additional depletion of real metals by factors of~2\\
                                                & low-metal -- high-metal\\
         3) Cosmic-ray ionization rate $\zeta$  & 1.3$\times10^{-18}$ s$^{-1}$ -- 1.3$\times10^{-16}$ s$^{-1}$\\
         4) Gas-density $n_H$		 	& 2$\times10^3$ cm$^{-3}$ -- 2$\times10^5$ cm$^{-3}$\\
         5) C$_{3}$H$_{2}$ + H fraction         &  0.50 - 1.00\\
         \noalign{\smallskip}
         \hline
      \end{array}
      \label{result-summary}
      $$
     % \begin{list}{}{}
	% \item  {\bf Note : {\sc Models 1-6} are constructed by varying each parameter above,
	%	combine all of them for another model, while the ``Default model" has default 
	%	conditions listed in Table~\ref{Table:default}.}
     % \end{list}
   \end{table}
   %%%%%%%%%%%%%%
   In our calculations of the $o/p$ abundance ratio of C$_{3}$H$_{2}$, we start with a
   default set of initial conditions, listed in Table \ref{Table:default},  used with
   our standard model of reactions.  The default model contains so-called low-metal
   abundances (depleted heavy elements with C/O = 0.41) with the additional 
   assumption that the  initial abundances are atomic except for hydrogen, which is
   molecular.  In addition, the total hydrogen density is set at 
   $n_{\rm H} = 2 \times 10^{4}$ cm$^{-3}$, the temperature at 10 K, the visual 
   extinction at 10 mag, and the cosmic-ray ionization rate 
   $\zeta$ at $1.3 \times 10^{-17}$ s$^{-1}$.  Finally the unknown branching fraction 
   for the dissociative recombination of C$_{3}$H$_{3}^{+}$ is set so that the 
   products C$_{3}$H$_{2}$ + H and C$_{3}$H + ...  are produced in equal amounts.  
   The chemistry is followed with a single-point model under constant physical 
   conditions.  

   In an attempt to determine how the computed $o/p$ abundance ratio for C$_{3}$H$_{2}$
   vs time depends upon an assortment of parameters, we have varied  the C/O 
   elemental ratio, the degree of metallicity, the cosmic ray ionization rate, and
   the gas density, as well as the neutral product branching ratio of the 
   dissociative recombination reaction of the ion $c$-C$_{3}$H$_{3}^{+}$.  These
   parameters affect how long the C$_{3}$H$_{2}$ ortho-para system can stay in the 
   loop shown in Fig. \ref{dominant-rxn}. Table~\ref{result-summary} shows the range
   of the parameters considered in this work for our standard list of additional reactions.  For each parameter, a number of different runs was attempted within the ranges listed in the table.
   The variation of parameters with our alternative model is more restrictive, as 
   will be discussed in Section \ref{SS:3.2}.   
%
%------------------------------------------------------------------------------
% 3.
\section{Results}
%
%------------------------------------------------------------------------------
%
   %---------------------------------------------------------------------------
   \subsection{Default and other standard sets of models vs observation} \label{SS:3.1}
\begin{figure*}
      \begin{center}
      \includegraphics[width=0.40\textwidth]{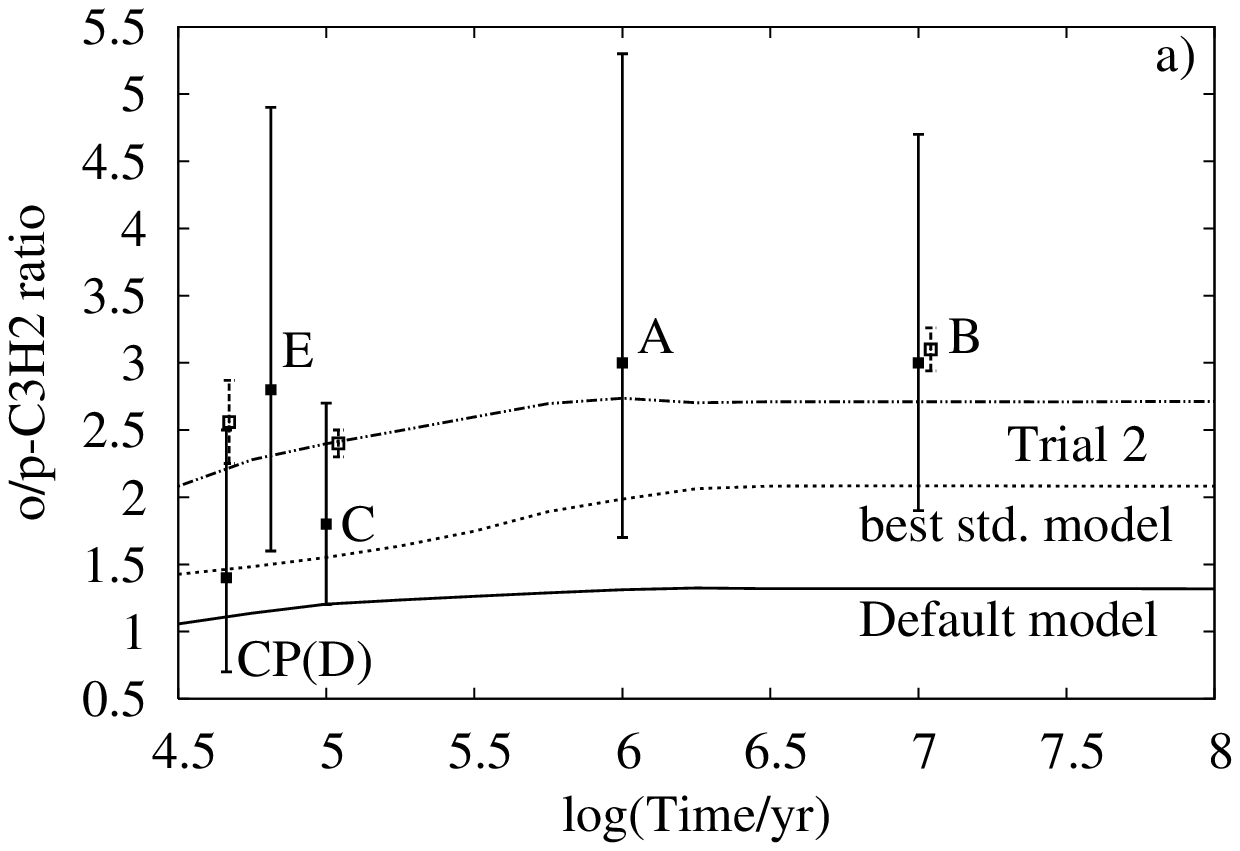}
      \includegraphics[width=0.40\textwidth]{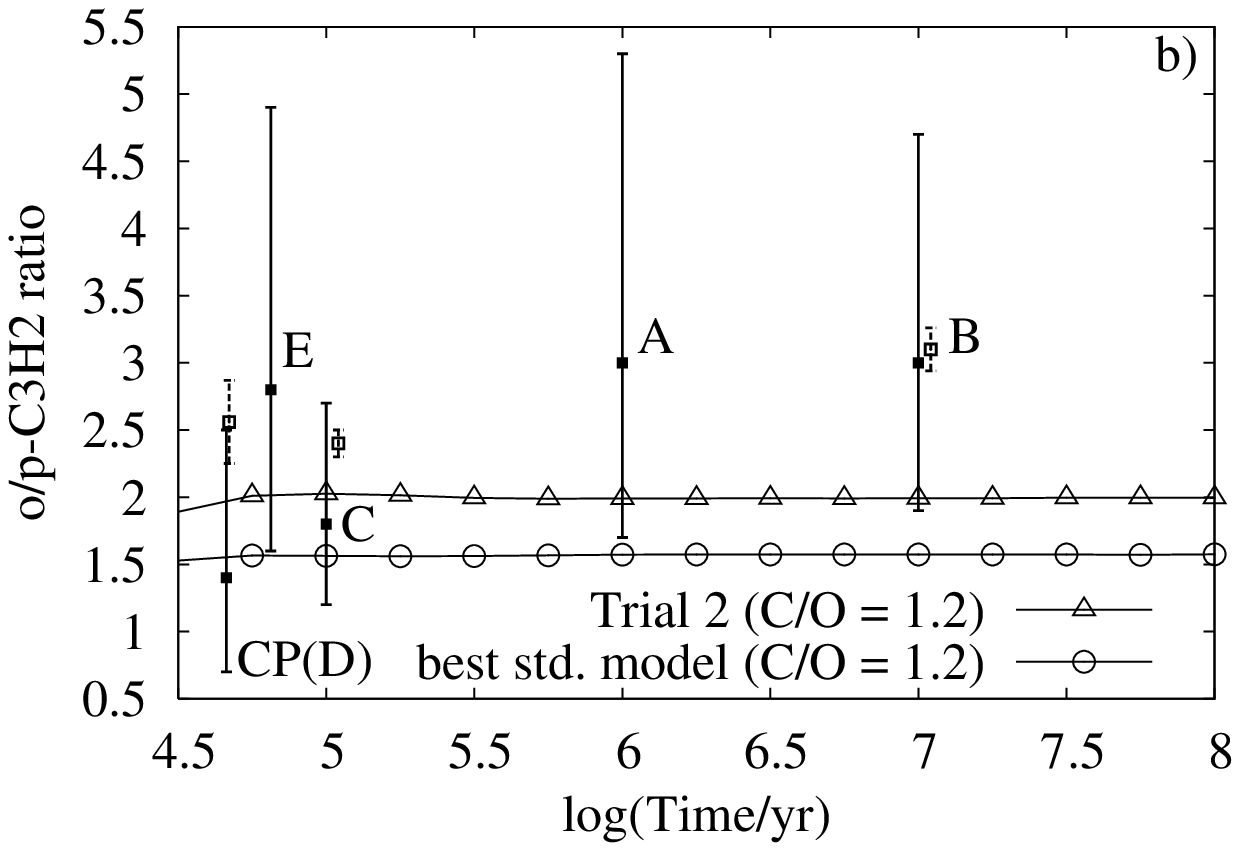}
      \includegraphics[width=0.40\textwidth]{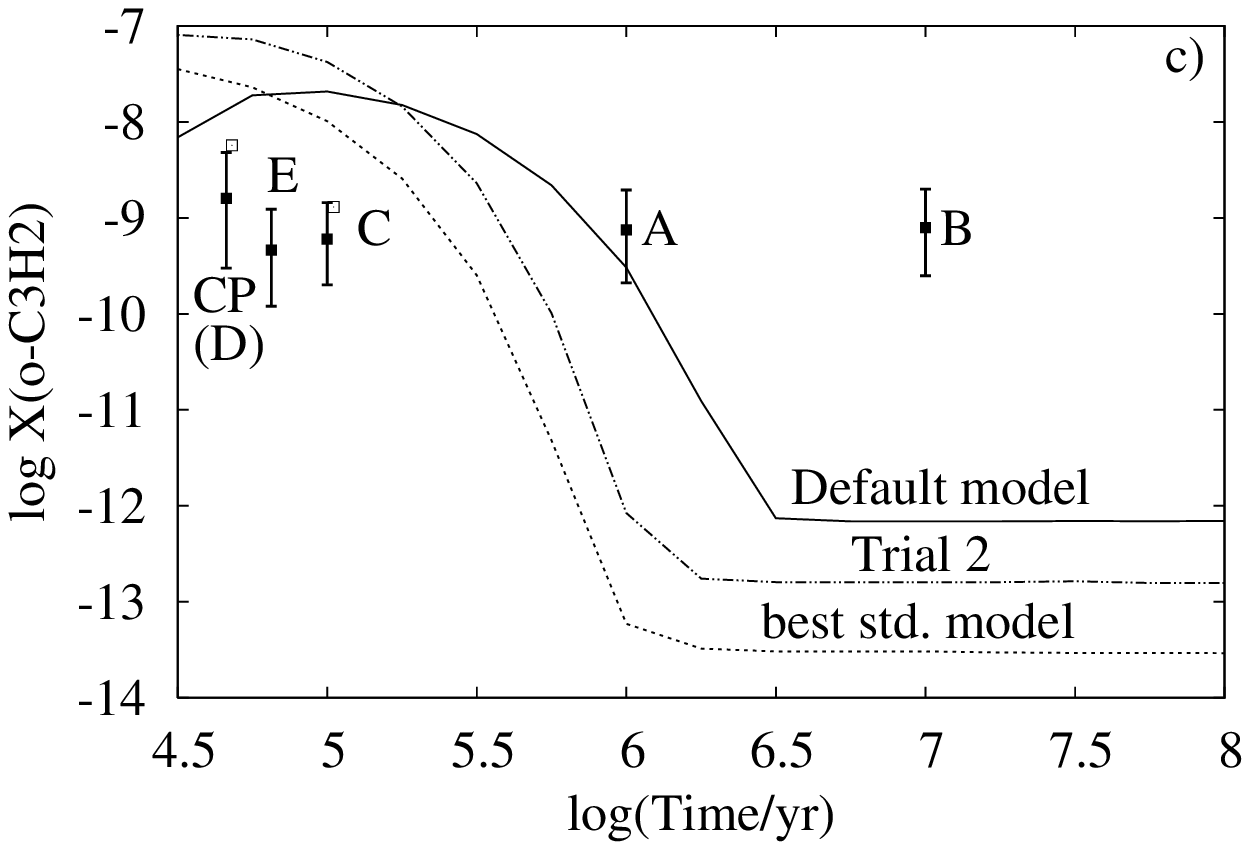}
      \includegraphics[width=0.40\textwidth]{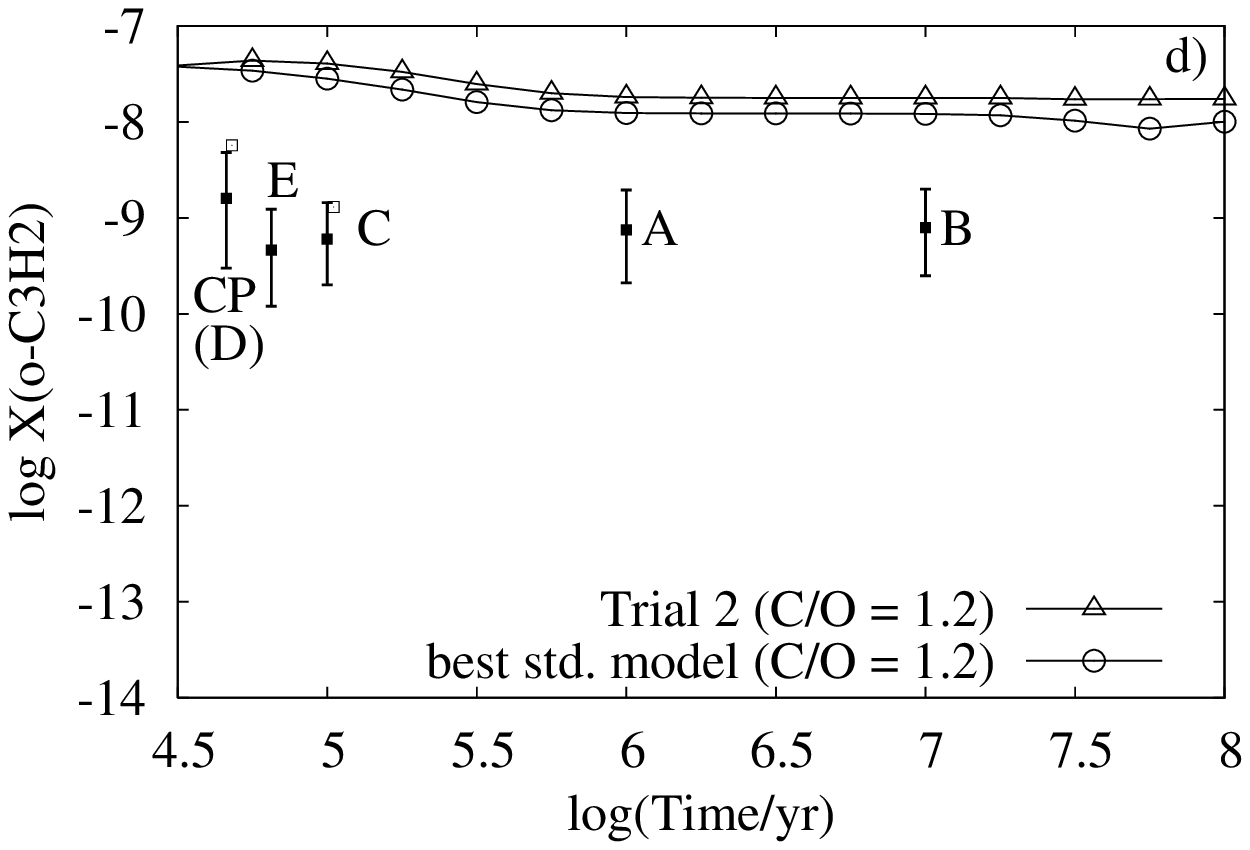}
      \end{center}
      \caption{Calculated values of the ortho-to-para C$_3$H$_2$ abundance ratio
         (panels a and b) and fractional abundance of $o$-C$_3$H$_2$ (panels c and d)
         with respect to H$_{2}$ vs time compared with observed values 
         in the cores of TMC-1. 
	 Note that the superimposed solid \citepalias{morisawa} and 
	 dotted \citep{madphd,cox-c3h2} error bars are from 
	 different observations.  See text for explanation of models used.}
   
      \label{op-results}
   \end{figure*}

   %---------------------------------------------------------------------------
   %
   The calculated $o/p$ ratio for C$_{3}$H$_{2}$ as a function of time in our default 
   model is shown in  panel a) of Fig. \ref{op-results}.  Superimposed are the 
   observational results for  the cores in TMC-1; e.g., Cores CP(D) and C, for which
   very short lifetimes are believed to pertain, Core A, with an intermediate
   lifetime, and Core B, for which a long lifetime of $\approx 10^{7}$ yr is normally 
   suggested.  Core E is included although its lifetime is rather uncertain: the high
   $o/p$ ratio suggests some evolution, but the measured NH$_{3}$/CCS ratio
   is low \citep{TMC-1-CP-b}, suggesting a young chemical age \citepalias{morisawa}.
   Uncertainties in the lifetimes are large and not included in the figure.
   
   With our default model, we obtain the result that the $o/p$ ratio is near unity at 
   times relevant to Cores CP(D) and C, in reasonable if not excellent agreement 
   with observed ratios,  but never increases very much with time, reaching a
   steady-state value under 1.5, in some disagreement with the measured ratios of 
   3.0($^{+2.2}_{-1.3}$) and  3.0($^{+1.7}_{-1.1}$) for Cores A and B, where we have used our recalculated uncertainties (see footnote in Table \ref{op-obs}).  The calculated fractional abundance 
   for $o$-C$_{3}$H$_{2}$ is shown in panel c) of Fig. \ref{op-results}, where it can
   be seen that the agreement with the lower age cores is fair, with the model 
   result being somewhat high, is excellent for Core A, but is particularly poor for 
   Core B. The calculated result is a typical one for pseudo-time-dependent models, 
   in which most organic species have high abundance only at so-called 
   ``early time.''  If the age of Core B were  10$^{6}$ yr, then the 
   disagreement would be eliminated for the fractional abundance of
   $o$-C$_{3}$H$_{2}$ although the disagreement with the observed $o/p$ ratio would 
   remain.    Using the abundance ratio of NH$_{3}$ to CCS as the criterion of
   chemical age, \citetalias{morisawa} considered the age of core B to be comparable to
   Core A, but this analysis is unclear because the abundance ratio is indeed larger
   for Core B, although not by a huge factor.  

   Given the probable disparity between the results of the default model and observations
   for the $o/p$ ratio at long times, we varied the five parameters discussed in Section 2.3 
   over the ranges listed in Table \ref{result-summary} to obtain a variety of
   results with the standard sets of models 1-6.  Note that the highest density used is similar to the densities derived by 
   \citetalias{morisawa}; our default value is the more traditional one.
   The result was little change in the quantities of interest except when we varied the branching ratio
   for the products of the dissociative recombination branching fraction.   
   When the branching fraction for the products C$_{3}$H$_{2}$ + H is increased to unity, 
   the calculated $o/p$ ratio for C$_{3}$H$_{2}$ increases to about 1.9 at steady state. 
   An increase in the ratio to about 2.1 can be obtained with the following other parameters:  
   C/O elemental ratio = 0.2, metal depletion a factor of two, no change in the value of $\zeta$, 
   and a higher density of $ 2 \times 10^{5}$ cm$^{-3}$.  The results of this model,
   labeled our  ``best'' standard model, for the $o/p$ ratio and abundance of 
   $o$-C$_{3}$H$_{2}$ are shown in panels a) and c) of Fig. \ref{op-results}. 
   The results for the $o/p$ ratio are now barely within the large observational error bars, 
   although the results for $o$-C$_{3}$H$_{2}$ are still poor for the evolved core B, 
   and would be poor even if the age of this core were reduced by an order of 
   magnitude.  As is well-known, the only way to increase the calculated abundance 
   of carbon-chain type species at late times is to use a carbon-rich elemental
   abundance.  With our best standard model changed only so that the elemental
   carbon-to-oxygen ratio C/O = 1.2, we get the results shown in panels b) and d)
   of Fig. \ref{op-results}, where it can be seen that the observed fractional 
   abundance of $o$-C$_{3}$H$_{2}$ as a function of core age is reasonably
   reproduced albeit somewhat high at all times with respect to the most recent
   measurements \citepalias{morisawa}, although the $o/p$ ratio has lost all of its time
   dependence.  Nevertheless, the model results for the $o/p$ ratio lie (barely) 
   within the observational errors except for Core B.   

   %
   %---------------------------------------------------------------------------
   \subsection{Models with direct interconversion}\label{SS:3.2} 
   %---------------------------------------------------------------------------
   %
   In order to reach an $o/p$ ratio of $\approx$ 3.0 at steady-state, we 
   included direct interconversion between the $para$ and $ortho$ forms of C$_{3}$H$_{2}$, 
   as shown in Table \ref{op-raxn}. For the alternative approach known as 
   ``Trial 2'', the conversion reactions are actually assumed to be more rapid than 
   protonation by a factor of five  for the protonating ions HX$^{+}$. 
   The results of Trial 2 using the same set of parameters as used for the best 
   standard model are also shown in panels a) and c) of Fig. \ref{op-results}.  
   Here the calculated ratio reaches a steady-state value of near 2.7 and is in good 
   agreement with observation over the complete time range. 
   On the other hand, the agreement with the fractional abundance of $o$-C$_{3}$H$_{2}$ 
   for late times (Core B) is once again poor, even if the age of this core is reduced 
   to 10$^{6}$ yr.  If, to remedy this latter deficiency, the C/O elemental ratio is
   set to 1.2, then the fractional abundance of $o$-C$_{3}$H$_{2}$ at late times is
   in fair agreement with the observed value in Core B but the calculated $o/p$ ratio
   never rises above 2.1.  
   If the less dramatic assumption is made that the rate coefficients for 
   interconversion are the same as for protonation (``Trial 1''), then the steady-state 
   $o/p$ ratio reaches a value of 2.5, also in reasonable agreement with observation 
   in the older cores.  But the same problems occur with the fractional abundance of
   $o$-C$_{3}$H$_{2}$ as occur for Trial 2.  Models with a C/O elemental ratio of
   1.0 are no more successful.  We conclude that no simple pseudo-time-dependent
   model of the type considered here explains all of the data.
   
      \begin{figure*}
      \centering
      \includegraphics[width=0.40\textwidth]{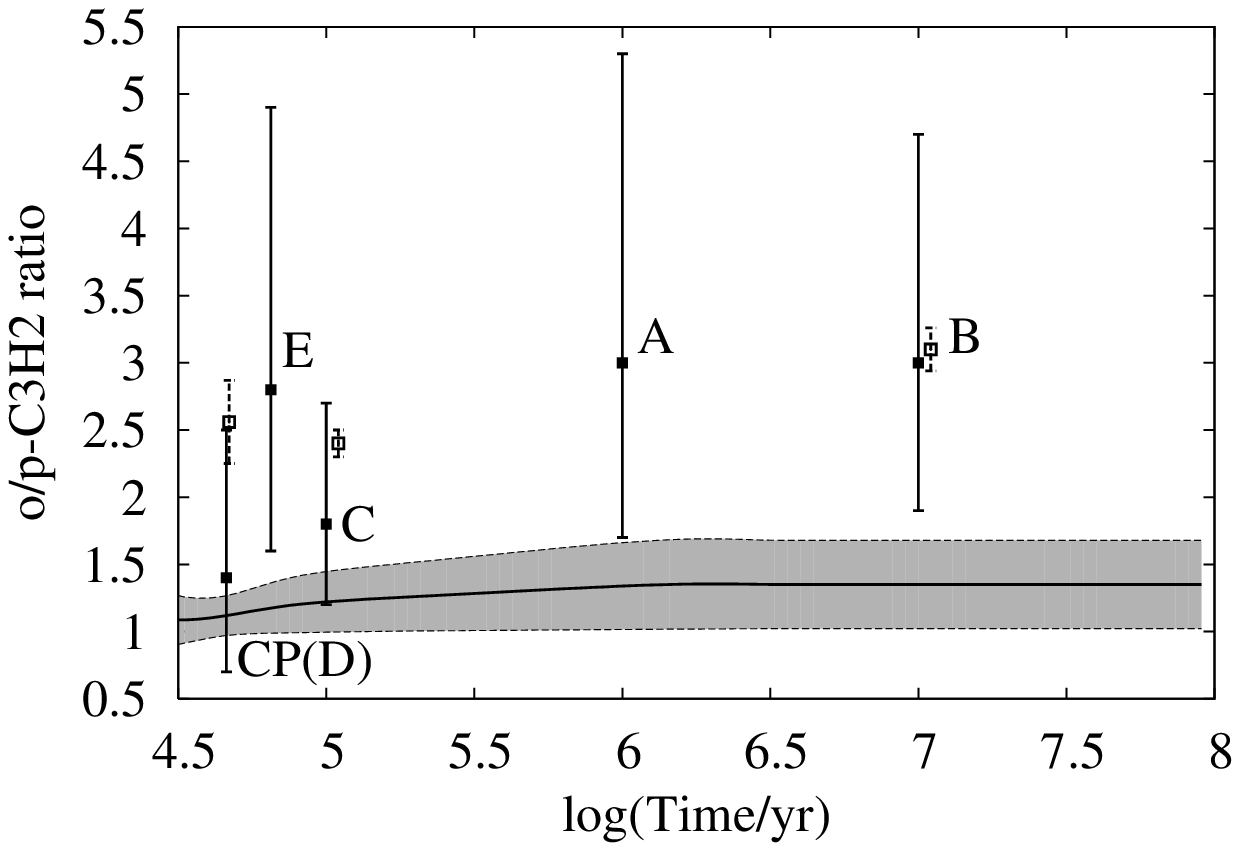}
      \includegraphics[width=0.40\textwidth]{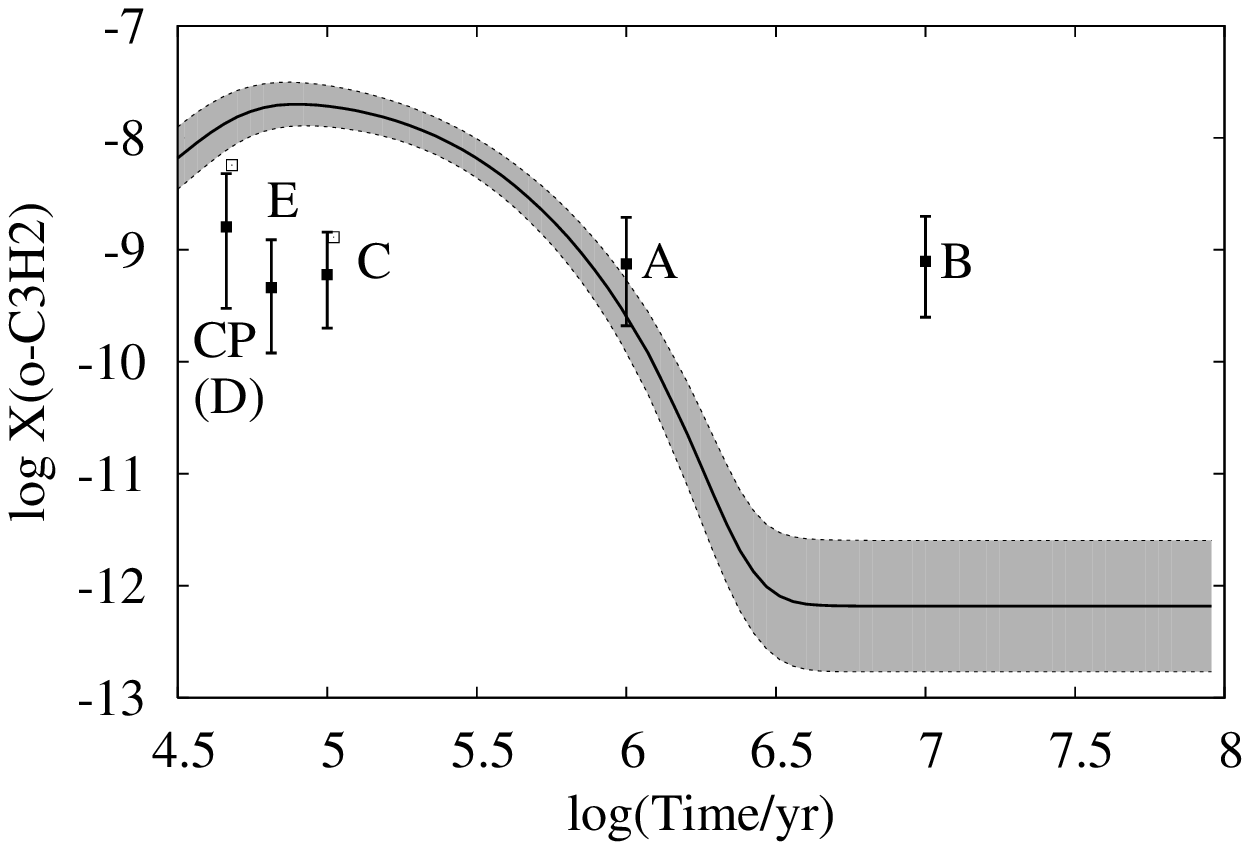}
      \caption{Calculated uncertainties in the ortho-to-para C$_3$H$_2$ abundance 
        ratio (left panel) and fractional abundance of $o$-C$_3$H$_2$ (right panel)
        for the default model case.}
      \label{op-error}
   \end{figure*}
   %
   %---------------------------------------------------------------------------
   \subsection{Comparison with simple model}
   %---------------------------------------------------------------------------
   %
   \citetalias{morisawa} developed a simple model to calculate the $o/p$ C$_{3}$H$_{2}$ 
   abundance ratio.  In their model only a few types of reactions are considered 
   (see $\bullet$ marks in Table \ref{op-raxn}).  In the limit of steady state, 
   the ratio can be shown to be 
   \begin{equation}
   \frac{
	{n(o\mbox{-C}_3\mbox{H}_2)}
	}
	{
	{n(p\mbox{-C}_3\mbox{H}_2)} 
	}
   = 
   \frac{
	(5+3\phi)\times \frac{
			     {n(o\mbox{-H}_2)}
			     }
			     {
		             {n(p\mbox{-H}_2)}
		 	     } + 3\phi + 3
	}
	{
	(1+\phi) \frac{
		      {n(o\mbox{-H}_2)}
		      }
		      {
		      {n(p\mbox{-H}_2)}
		      } + \phi + 3
	}
	\xrightarrow[]{o/p-H_2\sim0} \frac{3 (\phi + 1)}{\phi + 3}
   \end{equation}
   where  $\phi$ is $k_{DR}n(e) / k_{ion\mbox{-}Y} n(Y)$, 
   the ratio of the product of the dissociative recombination rate coefficient of 
   C$_{3}$H$_{3}^{+}$, $k_{DR}$, and the electron abundance, $n(e)$,
   divided by the product of the rate coefficient for ion-atom reactions of 
   C$_{3}$H$_{3}^{+}$, $k_{ion\mbox{-}Y}$, and the abundance of reactive atoms,
  $n(Y)$.  
   The parameter $\phi$ is essentially related to the ability of the reactants to 
   stay in the loop shown in Fig. \ref{dominant-rxn}.
   If one makes the assumption that H$_{2}$ is overwhelmingly $para$ in nature, then
   the $o/p$ ratio of C$_{3}$H$_{2}$ is determined solely by the parameter $\phi$,
   with a large value of $\phi$ leading to an $o/p$ ratio of 3 and a small value of 
   $\phi$ leading to a ratio of 1.  With the values for $n(Y)$ and $n(e)$ assumed by 
   \citetalias{morisawa}, the former ratio pertains at steady state; their assumptions 
   contain a rather high fractional abundance for electrons, which we would associate
   only with the high-metal case of elemental abundances.  With the more standard
   low-metal case, the computed ratio becomes very close to unity, in agreement with
   our more detailed results.  With the high-metal-type parameters chosen by
   \citetalias{morisawa}, their time-dependent calculation shows that the $o/p$ ratio of 
   C$_{3}$H$_{2}$ shifts dramatically from 1 to 3 at some intermediate time.  
   
   But there are two omissions in the theory leading to the simple steady-state 
   formula for the $o/p$ ratio: (i) destructive reactions involving C$_{3}$H$_{2}$ 
   with a host of non-protonating ions such as C$^{+}$ that make the system leave 
   the loop, and (ii) dissociative recombination leading to products other than 
   C$_{3}$H$_{2}$ + H, which again have the effect of forcing the system to leave
   the loop.  The effect of including both of these processes in the high-metal case makes the simple 
   steady-state for the $o/p$ ratio a more complex one and reduces the calculated 
   ratio to values close to those calculated in our more detailed approach. 
   In other words, it is still difficult to produce large $o/p$ abundance ratios in 
   the absence of direct inter-conversion reactions.

   %
   %---------------------------------------------------------------------------
   \subsection{The role of uncertainties in rate coefficients}
   %---------------------------------------------------------------------------
   %

   Another important aspect that we considered in our study is the uncertainty 
   in the rate coefficients. Using the method described in \citet{wakelam}, we 
   computed errors both in the fractional abundance of $o$-C$_3$H$_2$ and in 
   the $o/p$-C$_3$H$_2$ abundance ratio for the default model. Briefly, this 
   method consists of randomly generating new sets of reactions of the 
   {\sc osu.2003} database within their uncertainty range using a Monte-Carlo 
   procedure. We ran 2000 different runs which gave 2000 values of the  
   $o$-C$_3$H$_2$ abundance and of the $o/p$-C$_3$H$_2$ ratio. The error in the 
   abundance and ratio is  defined at each time as the envelope containing 
   66\% of the computed values (1$\sigma$). We chose a $1\sigma$ error 
   because the observations are also given an 
   observational error of $1\sigma$. We utilized an uncertainty of a factor of 
   two for the rate coefficients of the reactions listed in Table~\ref{op-raxn}. 
   For the reactions of this table that have more than one product channel, 
   we kept the branching ratio constant, because they were
   derived from theory and,  as a consequence, have no statistical error. 

    We found a maximum error of $\pm 0.4$ for the $o/p$-C$_3$H$_2$  ratio 
   at $10^4$~yr. This error decreases to $\pm 0.2$ at 10$^5$~yr before 
   increasing again to $\pm 0.3$ after 10$^6$~yr. As shown in Fig. \ref{op-error}, the 
   computed uncertainty in the ratio is not large enough to significantly affect any of
   our conclusions. The computed error in the fractional abundance of 
   $o$-C$_3$H$_2$ increases with time. Specifically, the error domain is 
   [X$_{o\mbox{-}C_3H_2}$/1.5, 1.5X$_{o\mbox{-}C_3H_2}$] at 
   10$^4$~yr and [X$_{o\mbox{-}C_3H_2}$/4, 4X$_{o\mbox{-}C_3H_2}$] at $10^7$~yr. 
   In this case again (see Fig. \ref{op-error}), the computed uncertainty 
   does not improve the agreement with observation significantly.  
%
%------------------------------------------------------------------------------
% 6.
\section{Discussion and Conclusion}
%
%------------------------------------------------------------------------------
%
   New observations by \citetalias{morisawa} of the ortho-to-para abundance ratio for 
   the cyclic isomer of C$_{3}$H$_{2}$ in the six cores of TMC-1 indicate that 
   this ratio is under 2.0 for two of three cores thought to be chemically
   young and is closer to the thermal value of 3.0 for those cores thought to be
   chemically more evolved.  The parameter used to determine chemical age is the 
   abundance ratio of NH$_{3}$ to that of CCS.  The $o/p$ ratio is determined by a
   chemistry in which the cyclic ion C$_{3}$H$_{3}^{+}$ is formed initially solely
   in its {\it para} state, but by a process of dissociative recombination and
   re-protonation, a significant ortho/para abundance ratio, near the thermal value 
   of 3.0,  can be built up given sufficient  time for high-metal abundances if competitive reactions are not 
   dominant.  Although a simple model by \citetalias{morisawa} indicates agreement with 
   their observations for five out of six cores if high-metal abundances are used, our more detailed calculations 
   with an extended version of the {\sc osu.2003} network of reactions do not give such a 
   definitive result, mainly because of competitive processes.  In fact, with 
   standard parameters varied over rather wide ranges, we find it difficult to 
   obtain $o/p$ ratios exceeding 2.0 at any time.  

   We are able to calculate higher 
   $o/p$ ratios for C$_{3}$H$_{2}$ only if we assume that the $ortho$ and $para$ forms
   of this species can be converted into one another by exchange processes involving
   H$^{+}$ and protonating ions HX$^{+}$.  This assumption is based on some 
   experiments involving H$_{3}^{+}$ by  \citet{cordon}   but here runs up against 
   the problem that both H$^{+}$ and HX$^{+}$ will likely react with C$_{3}$H$_{2}$
   and the competition between reaction and $ortho/para$ conversion may well favor 
   the former.  Thus, there is still no definitive chemical proof that gas-phase 
   processes can produce $ortho/para$ ratios for cyclic C$_{3}$H$_{2}$ near the thermal 
   value of 3.0.  If, then, the observational values in this range are sufficiently accurate to 
   distinguish correctly between low and high ratios,
   either the exchange reactions we consider between $o$-C$_{3}$H$_{2}$ and 
   $p$-C$_{3}$H$_{2}$ do occur and are actually favored over reaction, our branching fractions or other aspects of the gas-phase chemistry are in error, 
   or one must
   consider gas-grain interactions.  Future observations with a lower uncertainty will help to 
   determine which if any of these choices is the correct one.  Even if the $o/p$ ratio is reproducible by 
   pseudo-time-dependent models of the type considered here, we must realize that 
   the fractional abundance of the cyclic species tells a different story.  

%
%------------------------------------------------------------------------------
%
\begin{acknowledgements}
   We thank Prof. Momose's group (University of British Columbia) for providing 
   the TMC-1 observational data and their  manuscript prior to submission.  
   E.H., I.-H. P., and V.W. acknowledge the support of the National Science Foundation (US) 
   for support of our research program in astrochemistry.  We thank the Ohio Supercomputer 
   Center for computer time on their SV1 machine.
\end{acknowledgements}
%
%------------------------------------------------------------------------------
%

%%
%%------------------------------------------------------------------------------
\setcounter{table}{1}
 \begin{longtable}{lllcccc c cc}
  \caption[]{Dominant reactions, branching ratios, and rate coefficients for the C$_{3}$H$_{2}$ ortho/para problem}
   \label{op-raxn}\\

 %  \begin{tabular}{lllcccc c cc}
   \hline
   \noalign{\smallskip}
	                            &        &       & & \multicolumn{3}{c}{\sc Standard Models}		           & & \multicolumn{2}{c}{\sc Alternative Models}\\
	                            &        &       & & \multicolumn{3}{c}{-----------------------------------------------} 
                                                                                                                   & & \multicolumn{2}{c}{---------------------------------------}\\
                                    &        &       & & \multicolumn{3}{c}{\sc Default \& Sets of Models 1-6}                                  & & {\sc Trial1} $^a$ & {\sc Trial2} $^b$\\
   \noalign{\smallskip}
   \multicolumn{3}{c}{Reactions} 	             & BR $^c$     	& $\alpha$ $^d$ & $\beta$ $^d$  & SM $^e$ & & $\alpha$  & $\alpha$\\
   \noalign{\smallskip}
   \hline
   \noalign{\smallskip}
   \multicolumn{7}{l}{{\sc Formation}}\\
        C$_3$H$^+$   + $p$-H$_2$   & $\to$  & $p$-C$_3$H$_3^+$    	& $k_1$       	& 3.30(-13)  & 1.0    & $\bullet$ & & &\\
        $o$-C$_3$H$_3^+$ + e       & $\to$  & $o$-C$_3$H$_2$ + H  	& 1.0 $k_2$ 	& 3.15(-7)   & 0.5    & $\bullet$ & & &\\
                                   & $\to$  & $p$-C$_3$H$_2$ + H 	& 0.0  $k_2$  	& 0          & 0      & 	  & & &\\
                                   & $\to$  & C$_3$H + ......    	& $^f$ & 3.15(-7)   & 0.5    &           & & &\\
        $p$-C$_3$H$_3^+$ + e       & $\to$  & $o$-C$_3$H$_2$ + H	& 0.5 $k_2$ 	& 1.58(-7)   & 0.5    & $\bullet$ & & &\\
                                   & $\to$  & $p$-C$_3$H$_2$ + H 	& 0.5  $k_2$  	& 1.58(-7)   & 0.5    & $\bullet$ & & &\\
                                   & $\to$  & C$_3$H  + .....    	& $^f$	& 3.15(-7)   & 0.5    &           & & &\\
       $o$-C$_3$H$_2$ + H$_3^+$    & $\to$ & $o$-C$_3$H$_3^+$ + H$_2$  & 0.67$k_3$   & 5.16(-9)  & 0.5     & $\bullet$ & & 5.16(-9)  & 1/5 $\times$ 5.16(-9) \\ 
        	                   & $\to$ & $p$-C$_3$H$_3^+$ + H$_2$  & 0.33 $k_3$   & 2.54(-9)  & 0.5     & $\bullet$ & & 2.54(-9)  & 1/5 $\times$ 2.54(-9) \\ 
       $o$-C$_3$H$_2$ + H$_3$O$^+$ & $\to$ & $o$-C$_3$H$_3^+$ + H$_2$O & 0.67 $k_3'$  & 2.48(-9)  & 0.5     & 	  & & 2.48(-9)  & 1/5 $\times$ 2.48(-9) \\ 
        	                   & $\to$ & $p$-C$_3$H$_3^+$ + H$_2$O & 0.33 $k_3'$  & 1.22(-9)  & 0.5     & 	  & & 1.22(-9)  & 1/5 $\times$ 1.22(-9) \\ 
       $o$-C$_3$H$_2$ + HCO$^+$    & $\to$ & $o$-C$_3$H$_3^+$ + CO     & 0.67 $k_3''$ & 2.14(-9)  & 0.5     &  	  & & 2.14(-9)  & 1/5 $\times$ 2.14(-9) \\ 
         	                   & $\to$ & $p$-C$_3$H$_3^+$ + CO     & 0.33 $k_3''$ & 1.06(-9)  & 0.5     & 	  & & 1.06(-9)  & 1/5 $\times$ 1.06(-9) \\ 
   \noalign{\smallskip} 
   \noalign{\smallskip}
   \multicolumn{7}{l}{{\sc Destruction}}\\
      $p$-C$_3$H$_2$ + H$_3^+$    & $\to$ & $o$-C$_3$H$_3^+$ + H$_2$  & 0.0 $k_3$   & 0         & 0   & $\bullet$ & & 0        & 0        \\
       	                          & $\to$ & $p$-C$_3$H$_3^+$ + H$_2$  & 1.0 $k_3$   & 7.70(-9)  & 0.5 & $\bullet$ & & 7.70(-9) & 1/5 $\times$ 7.70(-9) \\
      $p$-C$_3$H$_2$ + H$_3$O$^+$ & $\to$ & $o$-C$_3$H$_3^+$ + H$_2$O & 0.0 $k_3'$  & 0         & 0   & 	       & & 0        & 0        \\
        	                  & $\to$ & $p$-C$_3$H$_3^+$ + H$_2$O & 1.0 $k_3'$  & 3.70(-9)  & 0.5 & 	       & & 3.70(-9) & 1/5 $\times$ 3.70(-9) \\
         $p$-C$_3$H$_2$ + HCO$^+$ & $\to$ & $o$-C$_3$H$_3^+$ + CO     & 0.0 $k_3''$ & 0         & 0   &           & & 0        & 0        \\
         	                  & $\to$ & $p$-C$_3$H$_3^+$ + CO     & 1.0 $k_3''$ & 3.20(-9)  & 0.5 &           & & 3.20(-9) & 1/5 $\times$ 3.20(-9) \\
   %%%
   \lbrack${o,\ p}$\rbrack-C$_3$H$_3^+$ + C  & $\to$ & products   & $k_4$    & 1.00(-9)  & 0 & $\bullet$  & & &\\
   \lbrack${o,\ p}$\rbrack-C$_3$H$_3^+$ + O  & $\to$ & products   & $k_4'$   & 4.50(-11) & 0 &            & & &\\
   \lbrack${o,\ p}$\rbrack-C$_3$H$_3^+$ + S  & $\to$ & products   & $k_4''$  & 1.00(-9)  & 0 &            & & &\\
   \lbrack${o,\ p}$\rbrack-C$_3$H$_3^+$ + Si & $\to$ & products   & $k_4'''$ & 2.00(-10) & 0 &            & & &\\
   %%%
   \lbrack$o,\ p$\rbrack-C$_3$H$_2$   + C$^+$   & $\to$ & products & $k_5$   & 4.20(-9)  & 0.5 & 	       & & &\\
   \lbrack$o,\ p$\rbrack-C$_3$H$_2$   + S$^+$   & $\to$ & products & $k_5'$  & 3.16(-9)  & 0.5 & 	       & & &\\
   \lbrack$o,\ p$\rbrack-C$_3$H$_2$   + Si$^+$  & $\to$ & products & $k_5''$ & 3.28(-9)  & 0.5 & 	       & & &\\
   \noalign{\smallskip}
   \noalign{\smallskip}
   \multicolumn{7}{l}{{\sc Interconversion by H$^{+}$}}\\
     $o$-C$_3$H$_2$   + H$^+$   & $\to$ & $p$-C$_3$H$_2$ + H$^+$   & $k_6$    &    & 0.5 &    & & 1/4 $\times$ 1.30(-8)&1/4 $\times$ 1.30(-8)  \\
     $p$-C$_3$H$_2$   + H$^+$   & $\to$ & $o$-C$_3$H$_2$ + H$^+$   & $k_{-6}$ &    & 0.5 &    & & 3/4 $\times$ 1.30(-8)& 3/4 $\times$ 1.30(-8) \\
   \noalign{\smallskip}
   \noalign{\smallskip}
   \multicolumn{7}{l}{{\sc Interconversion by HX$^+$}}\\
     $o$-C$_3$H$_2$ + H$_3^+$   & $\to$ & $p$-C$_3$H$_2$ + H$_3^+$    & $k_7$   & & 0.5&     & & 1/4 $\times$ 7.70(-9)  & 1/4 $\times$ 7.70(-9) \\ 
     $p$-C$_3$H$_2$ + H$_3^+$   & $\to$ & $o$-C$_3$H$_2$ + H$_3^+$    & $k_{-7}$   & & 0.5&     & & 3/4 $\times$ 7.70(-9) & 3/4 $\times$ 7.70(-9) \\
   \noalign{\smallskip}                                                                                              
    $o$-C$_3$H$_2$ + H$_3$O$^+$ & $\to$ & $p$-C$_3$H$_2$ + H$_3$O$^+$ & $k'_7$  & & 0.5&     & & 1/4 $\times$ 3.70(-9)  & 1/4 $\times$ 3.70(-9) \\
    $p$-C$_3$H$_2$ + H$_3$O$^+$ & $\to$ & $o$-C$_3$H$_2$ + H$_3$O$^+$ & $k'_{-7}$  & & 0.5&     & & 3/4 $\times$ 3.70(-9)  & 3/4 $\times$ 3.70(-9) \\
   \noalign{\smallskip}                                                                                              
       $o$-C$_3$H$_2$ + HCO$^+$ & $\to$ & $p$-C$_3$H$_2$ + HCO$^+$    & $k''_7$ & & 0.5&     & & 1/4 $\times$ 3.20(-9)  & 1/4 $\times$ 3.20(-9) \\
       $p$-C$_3$H$_2$ + HCO$^+$ & $\to$ & $o$-C$_3$H$_2$ + HCO$^+$    & $k''_{-7}$ & & 0.5&     & & 3/4 $\times$ 3.20(-9)  & 3/4 $\times$ 3.20(-9) \\
   \noalign{\smallskip}
   \hline
   %\end{tabular}
     \end{longtable}
      \begin{list}{}{}
	 \item  Note: $a(-b)$ = $a$ $\times$ 10$^{-b}$
	 \item  $^a$ The interconversion rates are obtained as discussed in Section \ref{SS:2.2}.
	 \item  $^b$ Competitive protonation rates are suppressed by factors of 5. 
	 \item  $^c$  Branching ratio between ortho and para modifications obtained
                by angular momentum algebra \citep{Ref:oka04}.
	 \item  $^d$ $k$ = $\alpha$\ (T/300)$^{-\beta}$ exp\ (-$\gamma$/T); $\alpha$ 
                is in units of cm$^{3}$ s$^{-1}$.
	 \item  $^e$ SM (Simple Model): the reactions marked with $\bullet$
	        are used in the model of \citetalias{morisawa};  
                with the generic forms HX$^+$ and Y. See Section 3.3.
	 \item  $^f$ In our standard approach, the branching fraction to form C$_{3}$H is set 
		equal to that for C$_{3}$H$_{2}$ 
	          except for models where it is  a free parameter.
      \end{list}

\end{document}